\documentclass[pdflatex,sn-mathphys,iicol]{sn-jnl}

\jyear{2025}%

\theoremstyle{thmstyleone}%
\theoremstyle{thmstyletwo}%

\theoremstyle{thmstylethree}%
\usepackage{soul,color}
\makeatletter
\AtBeginDocument{\let\hl\@firstofone}
\makeatother

\usepackage{subcaption}
\usepackage{graphicx}

\begin{document}

\title[Article Title]{Optimizing RAG Pipelines for Arabic: A Systematic Analysis of Core Components}


    
    



\author[1]{\fnm{Jumana} \sur{Alsubhi}}
\email{jumana.alsubhi@naseej.com}

\author[2]{\fnm{Mohammad D.} \sur{Alahmadi}}
\email{mdalahmadi@uj.edu.sa}

\author[1]{\fnm{Ahmed} \sur{Alhusayni}}
\email{ahmed.alhusainy@naseej.com}

\author[1]{\fnm{Ibrahim } \sur{Aldailami}}
\email{ebrahim.aldailami@naseej.com }

\author[1]{\fnm{Israa  } \sur{Hamdine}}
\email{israa.hamdin@naseej.com }

\author[1]{\fnm{Ahmad  } \sur{Shabana}}
\email{ashabana@naseej.com}

\author[1]{\fnm{Yazeed  } \sur{Iskandar}}
\email{yazeed.iskandar@naseej.com }

\author[1]{\fnm{Suhayb   } \sur{Khayyat}}
\email{souheib.khayyat@naseej.com }

\affil[1]{\orgname{Naseej Innovation Lab, Naseej for Technology, Riyadh}, \orgaddress{\country{Saudi Arabia}}}

\affil[2]{\orgdiv{Department of Software Engineering}, \orgname{College of Computer Science and Engineering, University of Jeddah}, \orgaddress{\city{Jeddah}, \postcode{23890}, \country{Saudi Arabia}}}



\abstract{Retrieval-Augmented Generation (RAG) has emerged as a powerful architecture for combining the precision of retrieval systems with the fluency of large language models. While several studies have investigated RAG pipelines for high-resource languages, the optimization of RAG components for Arabic remains underexplored. This study presents a comprehensive empirical evaluation of state-of-the-art RAG components—including chunking strategies, embedding models, rerankers, and language models—across a diverse set of Arabic datasets. Using the RAGAS framework, we systematically compare performance across four core metrics: context precision, context recall, answer faithfulness, and answer relevancy. Our experiments demonstrate that sentence-aware chunking outperforms all other segmentation methods, while BGE-M3 and Multilingual-E5-large emerge as the most effective embedding models. The inclusion of a reranker (bge-reranker-v2-m3) significantly boosts faithfulness in complex datasets, and Aya-8B surpasses StableLM in generation quality. These findings provide critical insights for building high-quality Arabic RAG pipelines and offer practical guidelines for selecting optimal components across different document types}

\keywords{
Large language models, retrieval-augmented generation, Arabic NLP, multilingual embeddings, chunking strategies, embedding models, and reranking.
}



\maketitle

\section{Introduction}

The rise of Retrieval-Augmented Generation (RAG) systems has significantly transformed how large language models (LLMs) access and utilize external knowledge sources \cite{gao2023retrieval}. By retrieving relevant information from large corpora before generating responses, RAG pipelines can produce more accurate, factual, and context-aware outputs \cite{lewis2020retrieval}. RAG is particularly valuable for Arabic, where challenges such as morphological richness and limited domain-specific resources can hinder the performance of standalone language models \cite{el2024exploring}.

Although recent advancements in RAG frameworks have led to significant progress in multilingual applications \cite{wu2024notall}, the systematic evaluation and optimization of RAG components for Arabic remain underexplored. Arabic's syntactic diversity, semantic density, and variable orthographic conventions introduce additional complexity across all stages of the RAG pipeline—from document chunking and representation to retrieval, reranking, and answer generation \cite{darwish2021panoramic,habash2010introduction}. Consequently, there is a need to investigate how Arabic-specific and multilingual models perform across these stages and how different pipeline configurations affect the quality of generated answers in Arabic.

This paper addresses this gap by presenting a comprehensive empirical evaluation of the key components in Arabic RAG pipelines. Our work involves controlled experimentation across six diverse Arabic datasets, evaluating four chunking strategies, six embedding models, one reranking model, and two large language models. We adopt the RAGAS framework to assess retrieval and generation quality along four dimensions: context precision, context recall, answer faithfulness, and answer relevancy. Our methodology is designed to reflect real-world applications of Arabic RAG systems in domains such as question answering, summarization, and knowledge-intensive dialogue.

To sum up, we make the following key contributions:

\noindent$\bullet$ We conduct a comprehensive empirical study of Retrieval-Augmented Generation (RAG) components—chunking strategies, embedding models, rerankers, and large language models—targeted specifically at Arabic language processing.\\

\noindent$\bullet$ We provide the first detailed comparison of chunking strategies in Arabic RAG pipelines, identifying when sentence-aware, semantic, recursive, or fixed-size chunking is most effective across diverse text types.\\

\noindent$\bullet$ We evaluate and benchmark six multilingual and Arabic-specific embedding models, offering insights into their effectiveness across different Arabic datasets and document structures.\\

\noindent$\bullet$ We analyze the impact of reranking on response accuracy and faithfulness, highlighting its benefits in complex and ambiguous retrieval scenarios.\\

\noindent$\bullet$ We assess the generation capabilities of two leading language models—Aya-8B and StableLM—demonstrating the superior performance of Aya-8B in Arabic answer generation within the RAG framework.\\

\noindent$\bullet$ We release a modular pipeline design and evaluation framework based on RAGAS, facilitating reproducible experimentation and future research in Arabic retrieval-augmented NLP systems.

Our experimental results show that sentence-aware chunking consistently outperforms other strategies in both context recall and answer relevancy, particularly in structured and semantically dense datasets. Among embedding models, \texttt{bge-m3} and \texttt{multilingual-e5-large} achieve the highest retrieval and generation scores across diverse domains. Moreover, applying a reranker substantially enhances faithfulness in complex retrieval settings, while Aya-8B demonstrates superior generation performance compared to StableLM in all evaluation scenarios.

The rest of the paper is organized as follows. In Section 2, we discuss related work. This is followed by Section 3, where we provide an overview of our empirical study. Section 4 is dedicated to presenting our results and key findings. The potential threats to the validity of our study are addressed in Section 5. Finally, we conclude our work in Section 6.

\section{Related Work}
Several studies have explored various aspects of RAG systems, though most focus on high-resource languages like English. Our work builds upon this foundation while addressing the specific challenges of Arabic language processing.

\subsection{Chunking Strategies}
Chunking strategies play a crucial role in determining retrieval effectiveness. Previous work by \cite{lewis2020retrieval} explored the interplay between chunking techniques and retrieval precision, demonstrating that fine-tuned chunking significantly improves performance. \cite{gao2023retrieval} investigated trade-offs between semantic chunking's computational complexity and its ability to maintain context integrity, emphasizing the importance of balancing computation and relevance. Additionally, \cite{chen2023retrieval} highlighted domain-specific chunking strategies, particularly for e-learning contexts, showing how retrieval strategies can be adapted to application-specific needs.

\subsection{Embedding Models}
Recent advances in multilingual embeddings have expanded the capabilities of RAG systems for non-English languages. The mGTE framework \cite{ni2022large} proposed a multilingual text retrieval approach supporting long-context understanding. BGE M3-Embedding \cite{chen2024bge} introduced a model designed to handle multiple languages, granularities, and tasks simultaneously. For Arabic specifically, \cite{abdelali2021pretraining} focused on improving Arabic NLP through nested embeddings that capture deeper semantic similarities. Moreover, \cite{el2024exploring} described efforts to create a RAG pipeline optimized for Arabic, covering all stages from document retrieval to generation.

\subsection{Language Models and Reranking}
The effectiveness of different language models in RAG pipelines has been examined by \cite{wu2024notall}, with particular attention to performance variations across languages and domains. Reranking mechanisms, which refine initial retrieval results before generation, have been shown to significantly improve RAG performance \cite{ma2023fine}, though their impact varies depending on the dataset and domain. For Arabic language models, \cite{antoun2020arabert} introduced AraBERT, a transformer-based model for Arabic language understanding that has been widely adopted in various NLP tasks. More recently, \cite{ousidhoum2024aya} presented Aya, a family of multilingual language models with strong performance on Arabic and other languages from the cohereforAI.

\subsection{Multilingual RAG Evaluation}
Evaluating RAG systems across languages presents unique challenges. \cite{thakur2024nomiracl} introduced NoMIRACL, a multilingual information retrieval benchmark covering 18 languages including Arabic. Similarly, \cite{wu2024notall} proposed Futurepedia, a benchmark for multilingual RAG that revealed significant performance disparities between high-resource and low-resource languages. The RAGAS framework \footnote{https://docs.ragas.io/en/stable/} has emerged as a standard for evaluating RAG systems, though its application to non-English languages requires careful consideration of language-specific factors. Our work extends these evaluation approaches by focusing specifically on Arabic, providing a comprehensive assessment of RAG components across diverse Arabic datasets.

\subsection{Arabic NLP Challenges}
Arabic presents unique challenges for NLP systems due to its morphological complexity and dialectal variations. \cite{darwish2021panoramic} provided a panoramic survey of NLP in the Arab world, highlighting resource limitations and technical challenges. \cite{habash2010introduction} detailed the linguistic properties that make Arabic processing particularly challenging, including its rich morphology and orthographic variations. These challenges extend to retrieval tasks, as demonstrated by \cite{abdelali2021pretraining}, who showed that standard tokenization and embedding approaches often underperform on Arabic text. Our work addresses these challenges by systematically evaluating RAG components optimized for Arabic language processing.

\begin{figure*}
\centering
    \includegraphics[width=0.9\textwidth]{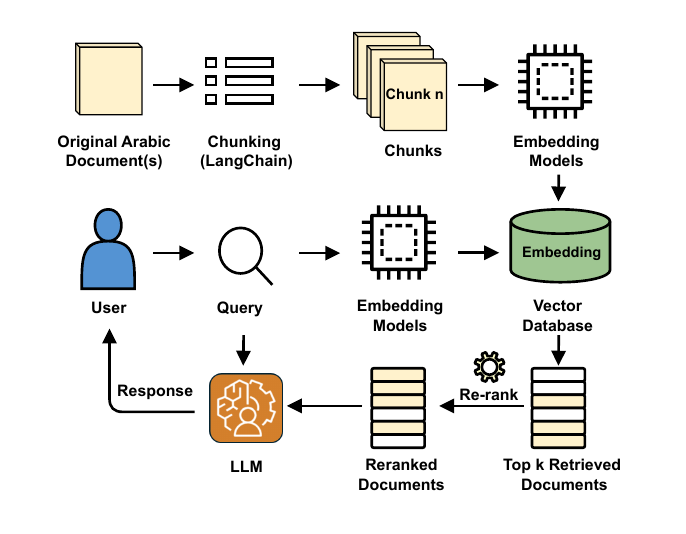}
    \vspace{-0.5em}
    \caption{Overview of the RAG pipeline used in this study, including chunking, embedding, retrieval, reranking, and answer generation.
    }\label{fig:overview}
    \vspace{-1.5em}
\end{figure*}

\section{Methodology}
In this section, we describe the datasets, pipeline design, and methodology used to evaluate the components of an end-to-end Retrieval-Augmented Generation (RAG) system for Arabic language tasks. Our goal is to empirically assess the effectiveness of various chunking strategies, embedding models, rerankers, and large language models (LLMs) across multiple Arabic datasets using the RAGAS evaluation framework.

\noindent\textbf{\textit{RQ$_1$}} \textit{What is the most effective chunking strategy for improving retrieval and generation quality in Arabic?}

\noindent\textbf{\textit{RQ$_2$}} \textit{How do multilingual and Arabic-specific embedding models compare in real-world RAG settings?}

\noindent\textbf{\textit{RQ$_3$}} \textit{What is the role of reranking in enhancing the faithfulness and relevance of generated answers?}

\noindent\textbf{\textit{RQ$_4$}} \textit{Which language models offer the best performance for Arabic generation in a RAG context?}

Figure \ref{fig:overview} provides a high-level summary of our experimental design. In the subsequent sections, we explain each component of the pipeline in detail as outlined in the figure.

\subsection{Datasets}

To ensure a comprehensive evaluation across a wide range of Arabic content types, linguistic complexities, and domain-specific characteristics, we selected six publicly available Arabic datasets. These datasets vary in terms of structure, semantic density, and intended task, providing a robust foundation for analyzing RAG pipeline behavior under different conditions.

\textbf{ARCD}~\cite{mozannar2019neural} is a high-quality Arabic reading comprehension dataset that contains factual passages paired with corresponding question-answer pairs, designed to evaluate machine comprehension in MSA (Modern Standard Arabic). \textbf{ArSQUAD}~\cite{arkady2019squad} is an Arabic adaptation of the popular English SQuAD dataset, featuring extractive QA tasks that enable benchmarking in information-seeking contexts. \textbf{SaudiWiki}~\cite{alqahtani2022saudiwiki} is a large-scale Arabic corpus derived from Wikipedia articles specific to Saudi Arabia, offering a rich source of factual, encyclopedic content with varied topical coverage.

To evaluate inference and reasoning abilities, we included \textbf{QA4MRE}~\cite{penas2013qa4mre}, a multilingual benchmark dataset containing complex, multi-paragraph texts and challenging reading comprehension questions. We also utilized \textbf{Quran Tafseer}~\cite{altammami2020automatic}, which comprises semantically rich religious texts and interpretive explanations that require deep contextual understanding. Lastly, the \textbf{Hindawi Books} collection~\cite{zerrouki2012hindawi} provides literary Arabic content with diverse writing styles and unstructured narrative formats, allowing us to assess RAG component behavior on long-form, less formalized text.

Together, these datasets facilitate a rigorous and multifaceted evaluation of chunking, retrieval, reranking, and generation strategies in Arabic RAG pipelines.

\subsection{Chunking Strategies}
\label{subsec:chunking}

Chunking is a critical preprocessing step in Retrieval-Augmented Generation (RAG) systems, as it directly impacts the granularity and coherence of retrievable information. In this study, we evaluate four distinct chunking strategies to determine their effect on retrieval precision, context relevance, and downstream generation quality in Arabic-language pipelines. Each strategy was implemented using the LangChain\footnote{https://www.langchain.com/} framework to ensure consistency and modularity across experimental runs.

\textbf{Fixed-size chunking} segments the input text into equal-length chunks based on a predefined character or token limit. While this approach is computationally efficient and simple to implement, it often disregards semantic boundaries, potentially splitting coherent sentences or logical units across chunks.

\textbf{Recursive chunking} employs a hierarchical splitting mechanism that recursively applies a set of fallback delimiters—such as paragraph breaks, sentence boundaries, and word-level tokens—to ensure that each chunk respects natural textual boundaries while remaining within the specified size constraints. This method seeks to balance structure preservation and chunk uniformity.

\textbf{Sentence-aware chunking} prioritizes semantic integrity by aligning chunk boundaries with sentence delimiters. By preserving complete sentences within each chunk, this strategy is designed to minimize context fragmentation, which is particularly valuable in tasks requiring deep linguistic understanding or reasoning.

\textbf{Semantic chunking} utilizes embedding-based similarity measures to group semantically related sentences into cohesive units. While this method offers the potential to maximize context relevance, it is the most computationally intensive due to the need for pairwise sentence comparisons and clustering. Moreover, its performance may vary significantly depending on the quality of the underlying embedding model.

Through systematic experimentation across six Arabic datasets, we evaluate the impact of these chunking strategies on RAG pipeline performance using RAGAS metrics. This analysis provides insights into the trade-offs between computational efficiency and semantic coherence when selecting chunking methods for Arabic text processing.

\subsection{Embedding Models}
\label{subsec:embedding_models}

Embedding models play a central role in Retrieval-Augmented Generation (RAG) systems by converting textual chunks and user queries into dense vector representations that facilitate similarity-based retrieval. The quality of these embeddings directly influences the relevance of retrieved contexts, which in turn affects the overall accuracy and faithfulness of the generated responses. Given the unique linguistic characteristics of Arabic—including rich morphology, script variability, and limited high-resource training data—we evaluate both Arabic-specific and multilingual embedding models.

Our study includes six models selected for their relevance to Arabic NLP tasks and prior performance in multilingual retrieval benchmarks. These are: bge-m3 by BAAI, multilingual-e5-large by Microsoft, snowflake-arctic-embed-l-v2.0 by Snowflake, gte-multilingual-base by Alibaba, Arabic-Triplet-Matryoshka-V2, and Arabic-mpnet-base-all-nli-triplet by Omartificial Intelligence. The selected models vary in size, training objectives, and support for Arabic, providing a diverse comparison set.

BGE-m3 and multilingual-e5-large are both high-capacity multilingual embedding models trained on cross-lingual retrieval tasks using advanced contrastive learning techniques. These models are designed for general-purpose semantic search and have demonstrated strong performance across diverse languages. Snowflake-arctic-embed-l-v2.0 and GTE-multilingual-base offer efficient alternatives with competitive performance, particularly in low-latency scenarios. The two Arabic-specific models—Arabic-Triplet-Matryoshka-V2 and Arabic-mpnet-base-all-nli-triplet— are optimized for Arabic language understanding and were included to assess whether specialized training leads to improved performance in Arabic-only contexts.

For consistency, all embeddings were generated using sentence-aware chunking, as this strategy demonstrated superior performance in our earlier experiments. Embedding vectors were indexed using FAISS and retrieved based on cosine similarity. The performance of each embedding model was evaluated using the RAGAS framework across six Arabic datasets, focusing on context precision, recall, and their impact on answer generation quality.

\subsection{Reranking Component}
While initial retrieval based on embedding similarity often yields contextually relevant passages, it may not always prioritize the most factually aligned or semantically coherent chunks with respect to the input query. To address this, reranking modules are commonly introduced to refine the order of retrieved results prior to generation. In Retrieval-Augmented Generation (RAG) systems, rerankers can significantly enhance downstream faithfulness and precision by promoting more relevant contexts and demoting spurious or ambiguous ones.

In our study, we incorporate the \texttt{bge-reranker-v2-m3} model, a multilingual reranker optimized for pairwise relevance scoring using dense representations. This model has demonstrated strong performance across various reranking benchmarks and supports Arabic input natively. After retrieving the top-$k$ candidate contexts via cosine similarity from the embedding index, the reranker is applied to reorder them based on their alignment with the user query.

To isolate the impact of reranking, we conducted paired experiments with and without the reranker across all datasets while keeping other pipeline components constant. The effect of reranking was assessed using RAGAS metrics, particularly focusing on improvements in answer faithfulness and overall generation quality. Our analysis also explores dataset-level variations, where reranking tends to yield greater benefits in complex, ambiguous, or loosely structured domains such as QA4MRE and Hindawi Books, while having marginal effect in highly structured sources like SaudiWiki.

\subsection{Language Models}

The final component in the Retrieval-Augmented Generation (RAG) pipeline is the language model responsible for synthesizing an answer based on the retrieved context. The performance of the language model is crucial, as it determines not only the fluency and coherence of the response but also its factual alignment with the retrieved evidence. This is particularly important in Arabic, where morphological richness and syntactic flexibility increase the challenge of generating precise and contextually appropriate text.

We evaluate two open-source generative models in this study: \texttt{Aya-8B} and \texttt{StableLM-1.6B}. \texttt{Aya-8B} is a large-scale Arabic-focused language model developed to support high-quality text generation in Modern Standard Arabic and dialectal variants. Its training data includes a mixture of Arabic web content, Wikipedia, and curated question answering corpora, making it well-suited for Arabic NLP tasks. In contrast, \texttt{StableLM-1.6B} is a general-purpose multilingual model with smaller capacity, designed for efficient deployment across a wide range of languages.

Both models were evaluated under identical conditions using the same retrieved contexts (with and without reranking), chunking strategy (sentence-aware), and embedding model (\texttt{bge-m3}). We measured performance using RAGAS metrics, with particular attention to answer faithfulness and relevancy. Our analysis highlights the comparative advantages of \texttt{Aya-8B} in capturing semantic nuance and maintaining factual consistency, especially in datasets with high linguistic complexity such as Quran Tafseer and QA4MRE.

\subsection{Evaluation Metrics}
\label{subsec:evaluation_metrics}

To evaluate the performance of the Retrieval-Augmented Generation (RAG) pipeline and its components, we adopt the RAGAS framework~\cite{es2023ragas}, a standardized metric suite specifically designed to assess retrieval-augmented systems in a unified and interpretable manner. RAGAS provides a modular evaluation approach that decomposes performance into retrieval quality, generation quality, and their interplay.

We use four core metrics from RAGAS in this study:

\textbf{Context Precision} measures the proportion of retrieved context that is relevant to the query. High precision indicates that the retriever effectively filters out irrelevant or noisy passages.

\textbf{Context Recall} evaluates whether all necessary information required to generate a correct response has been retrieved. It captures the system's ability to cover all supporting evidence, which is particularly important in reasoning-intensive tasks.

\textbf{Faithfulness} assesses whether the generated answer is factually supported by the retrieved context. This metric is critical for minimizing hallucinations and ensuring that the language model does not generate content inconsistent with the source material.

\textbf{Answer Relevancy} measures the degree to which the generated answer aligns with the intent of the user query. It reflects both topical alignment and linguistic appropriateness.

All evaluations were conducted using manually curated ground-truth answers where available. For datasets without explicit answer labels, we employed semi-automated evaluation scripts with expert validation to ensure quality. The use of these metrics enables a fine-grained understanding of how each RAG component contributes to overall system performance, facilitating component-wise analysis and cross-dataset comparisons.

\subsection{Results}
\subsection{RQ\textsubscript{1}: Chunking Strategy}

\noindent \textbf{Motivation:} Chunking is a foundational preprocessing step in Retrieval-Augmented Generation (RAG) systems. The way documents are segmented can significantly affect the quality of retrieval and the contextual integrity of generated answers. Given the structural and syntactic diversity across Arabic texts, we aim to identify which chunking strategy best balances semantic coherence and retrieval efficiency across different datasets and task types.

\noindent \textbf{Findings and Analysis:} Our results show that \textit{sentence-aware chunking} achieves the highest average overall score across all datasets (\textbf{74.78}), outperforming fixed-size (\textbf{69.41}), recursive (\textbf{69.13}), and semantic chunking (\textbf{66.92}). Sentence-aware chunking leads in four out of six datasets: \textbf{ARCD} (75.12), \textbf{QA4MRE} (80.50), \textbf{Hindawi Books} (76.49), and shows competitive performance in \textbf{Quran Tafseer} (81.63), only slightly behind semantic chunking (82.45). Fixed-size chunking performs best on the \textbf{SaudiWiki} dataset (86.69), likely due to the structured and uniform nature of Wikipedia articles. Recursive chunking performs comparably in \textbf{ArSQUAD}, leading slightly with a score of 52.35 compared to sentence-aware's 52.21, but both approaches remain significantly higher than semantic chunking for that dataset. Interestingly, \textit{semantic chunking}, while theoretically advantageous due to its embedding-based grouping of related sentences, consistently underperforms across all datasets, with particularly poor results on \textbf{ARCD} (74.59), \textbf{QA4MRE} (50.98), and \textbf{SaudiWiki} (72.06). This suggests that current semantic chunking implementations may introduce noise or disrupt coherence, especially in morphologically complex Arabic text. Overall, sentence-aware chunking appears to offer the best trade-off between semantic integrity and chunk length control, contributing to higher faithfulness and answer relevancy scores in both structured and unstructured domains.

\textbf{Summary Answer to RQ\textsubscript{1}:} Sentence-aware chunking outperformed all other strategies in overall RAG performance, providing the most effective balance between semantic integrity and retrieval precision. Its consistency across diverse datasets—especially those requiring reasoning or narrative comprehension—highlights its suitability for Arabic text segmentation in RAG pipelines.

\begin{table*}[t]
    \centering
    \caption{Comparison of chunking strategies across six Arabic datasets based on overall mean scores. The best-performing score per dataset is shown in \textbf{bold}.}

        \begin{tabular}{l|cccc}
            \hline
            \textbf{Dataset} & \textbf{Fixed-size} & \textbf{Recursive} & \textbf{Sentence-aware} & \textbf{Semantic} \\
            \hline
            ARCD           & 74.74 & 72.82 & \textbf{75.12} & 74.59 \\
            ArSQUAD        & 51.12 & \textbf{52.35} & 52.21 & 49.64 \\
            SaudiWiki      & \textbf{86.69} & 85.27 & 82.76 & 72.06 \\
            QA4MRE         & 50.25 & 47.89 & \textbf{80.50} & 50.98 \\
            Quran Tafseer  & 80.20 & 81.43 & 81.63 & \textbf{82.45} \\
            Hindawi Books  & 73.46 & 75.04 & \textbf{76.49} & 71.79 \\
            \hline
            \textbf{Average} & 69.41 & 69.13 & \textbf{74.78} & 66.92 \\
            \hline
        \end{tabular}
    
\end{table*}

\subsection{RQ\textsubscript{2}: Embedding Model}
\label{subsec:embedding_results}

\noindent \textbf{Motivation:} Embedding models are critical to the RAG pipeline, as they encode both queries and document chunks into dense vector representations used for retrieval. In Arabic, where semantic similarity can be obfuscated by rich morphology and syntactic variance, the choice of embedding model significantly impacts retrieval effectiveness and downstream generation quality.

\noindent \textbf{Findings and Analysis:} Our results indicate that \texttt{bge-m3} achieved the highest overall score (\textbf{70.99}) across the six datasets, closely followed by \texttt{multilingual-e5-large} (\textbf{70.31}). These two models consistently outperformed others in both retrieval precision and answer generation metrics. Notably, \texttt{bge-m3} excelled in datasets requiring complex reasoning such as \textbf{ARCD} (80.29) and \textbf{Quran Tafseer} (82.72), while \texttt{multilingual-e5-large} performed strongly in structured corpora like \textbf{SaudiWiki} (88.74). Among the Arabic-specific models, \texttt{Arabic-Triplet-Matryoshka-V2} showed promising results with an overall average of 66.46, especially in \textbf{Quran Tafseer} and \textbf{Hindawi Books}, where the dataset language closely matched its training domain. However, both \texttt{Arabic-mpnet-base-all-nli-triplet} and \texttt{snowflake-arctic-embed-l-v2.0} underperformed, with average scores of 45.92 and 69.48, respectively, reflecting limited generalizability or weaker adaptation to complex retrieval tasks. Multilingual models showed more balanced performance across datasets, with \texttt{gte-multilingual-base} achieving a respectable average (68.48), particularly effective in mid-scale structured domains like \textbf{Hindawi Books}. The performance gap between the top two models and others highlights the importance of strong cross-lingual training and large model capacity for high-recall retrieval in Arabic. Overall, our analysis suggests that multilingual, contrastively trained models such as \texttt{bge-m3} and \texttt{multilingual-e5-large} provide superior retrieval performance in Arabic RAG pipelines, even compared to Arabic-specialized models.

\textbf{Summary Answer to RQ\textsubscript{2}:} Among the evaluated models, \texttt{bge-m3} and \texttt{multilingual-e5-large} delivered the strongest and most consistent performance across all datasets. Their multilingual, contrastively trained embeddings outperformed Arabic-specific models, confirming the effectiveness of large-scale, general-purpose embeddings for Arabic retrieval tasks in RAG pipelines.

\begin{table*}[t]
    \centering
    \caption{Overall RAG performance scores across datasets using different embedding models. The highest score per dataset is highlighted in \textbf{bold}.}
    \label{tab:embedding_results}
    \resizebox{\textwidth}{!}{
        \begin{tabular}{l|cccccc}
            \hline
            \textbf{Dataset} & \textbf{Arctic-Embed} & \textbf{Arabic-MPNet} & \textbf{Matryoshka} & \textbf{GTE-Base} & \textbf{BGE-M3} & \textbf{E5-Large} \\
            \hline
            ARCD           & 78.99  & 55.49  & 79.09  & 76.94  & \textbf{80.29} & 78.43 \\
            ArSQUAD        & 48.39  & 28.66  & 45.55  & 49.54  & 51.97  & \textbf{56.55} \\
            SaudiWiki      & 85.21  & 58.21  & 83.96  & 86.87  & 88.31  & \textbf{88.74} \\
            QA4MRE         & \textbf{50.57}  & 34.20  & 46.54  & 49.64  & 48.97  & 49.16 \\
            Quran Tafseer  & 79.40  & 53.86  & 75.53  & 77.68  & \textbf{82.72} & 81.16 \\
            Hindawi        & \textbf{74.34}  & 45.11  & 68.15  & 70.25  & 73.70  & 67.84 \\
            \hline
            \textbf{Average} & 69.48 & 45.92 & 66.46 & 68.48 & \textbf{70.99} & 70.31 \\
            \hline
        \end{tabular}
    }
\end{table*}

\subsection{RQ\textsubscript{3}: Reranking Impact}
\label{subsec:reranking_results}

\noindent \textbf{Motivation:} Reranking aims to improve retrieval quality by refining the ordering of the top-$k$ retrieved chunks based on their semantic alignment with the query. In RAG systems, especially for Arabic where lexical variance is high, reranking can help prioritize the most contextually relevant passages, thus enhancing the factual consistency and relevance of the generated answers.

\noindent \textbf{Findings and Analysis:} Integrating the \texttt{bge-reranker-v2-m3} module led to noticeable improvements across most datasets, increasing the overall average RAG score from \textbf{70.99} (without reranker) to \textbf{74.15}. The reranker was particularly impactful in datasets requiring deeper semantic matching or complex reasoning. The largest improvement was observed in \textbf{ARCD}, where the score rose from 80.29 to \textbf{86.31}, driven by significant gains in context precision and faithfulness (the latter improving by over 15 points). Similarly, \textbf{QA4MRE} saw a 4.6-point increase in overall score, highlighting the reranker’s ability to promote more useful chunks in inference-heavy tasks. \textbf{ArSQUAD} and \textbf{Hindawi Books} also benefited, showing gains of 4.49 and 4.04 points respectively, particularly in faithfulness and answer relevancy. In contrast, the impact was mixed in structured datasets. For instance, \textbf{SaudiWiki} saw a slight drop in overall score despite improvements in precision, suggesting that reranking may offer limited benefits when the initial retrieval quality is already high. Similarly, \textbf{Quran Tafseer} saw only a modest improvement of 0.73 points. Overall, reranking is most beneficial in settings where retrieved chunks are noisy or semantically ambiguous. Its effectiveness is particularly evident in datasets that require contextual inference, making it a valuable addition to Arabic RAG pipelines where document structure varies widely.

\textbf{Summary Answer to RQ\textsubscript{3}:} The inclusion of the \texttt{bge-reranker-v2-m3} significantly improved overall RAG scores—particularly in complex or ambiguous datasets like ARCD and QA4MRE—by enhancing faithfulness and answer relevancy. Its impact was less pronounced in highly structured datasets, suggesting reranking is most beneficial when initial retrieval lacks semantic precision.

\begin{table}[ht]
\centering
\caption{Comparison of RAG performance with and without the \texttt{bge-reranker-v2-m3} reranker. Values represent overall mean scores across RAGAS metrics.}
\label{tab:reranking_results}
\resizebox{\linewidth}{!}{
\begin{tabular}{l|cc}
\hline
\textbf{Dataset} & \textbf{Without Reranker} & \textbf{With Reranker} \\
\hline
ARCD           & 80.29  & \textbf{86.31} \\
ArSQUAD        & 51.97  & \textbf{56.47} \\
SaudiWiki      & \textbf{88.31}  & 87.39 \\
QA4MRE         & 48.97  & \textbf{53.57} \\
Quran Tafseer  & 82.72  & \textbf{83.45} \\
Hindawi Books  & 73.70  & \textbf{77.74} \\
\hline
\textbf{Average} & 70.99 & \textbf{74.15} \\
\hline
\end{tabular}
}
\end{table}

\subsection{RQ\textsubscript{4}: Language Model Performance}

\label{subsec:llm_results}

\noindent \textbf{Motivation:} In Retrieval-Augmented Generation (RAG) pipelines, the language model is responsible for synthesizing answers based on the retrieved context. The effectiveness of this stage hinges not only on fluency and coherence, but also on the model’s ability to generate answers that are faithful, relevant, and semantically aligned with both the query and supporting evidence. For Arabic, evaluating the generation capabilities of large language models (LLMs) is especially important due to syntactic variability and limited high-quality generation resources.

\noindent \textbf{Findings and Analysis:} We compared the performance of two models—\texttt{Aya-8B} and \texttt{StableLM-1.6B}—using consistent chunking (sentence-aware), embeddings (\texttt{bge-m3}), and retrieval configurations. The results show that \texttt{Aya-8B} consistently outperforms \texttt{StableLM} across all datasets, achieving a higher overall average score (\textbf{72.06} vs. \textbf{67.08}). The performance gap is most notable in inference-heavy datasets such as \textbf{QA4MRE}, where \texttt{Aya-8B} achieved a score of 61.8, compared to only 55.07 with \texttt{StableLM}. Similarly, in \textbf{Quran Tafseer}, where semantic precision is crucial, \texttt{Aya-8B} reached 78.38 versus 71.24 for \texttt{StableLM}. In \textbf{SaudiWiki}, a structured factual dataset, \texttt{Aya-8B} achieved the highest overall score of  88.81. These results highlight the benefit of domain-tuned, Arabic-capable models for accurate and relevant answer generation. While \texttt{StableLM} demonstrates reasonable performance in simpler tasks, it struggles with deep contextual inference, often resulting in less faithful or semantically off-target responses. Overall, \texttt{Aya-8B} proves to be a more robust choice for Arabic RAG pipelines, especially in high-stakes or complex reasoning scenarios.

\textbf{Summary Answer to RQ\textsubscript{4}:} \texttt{Aya-8B} consistently outperformed \texttt{StableLM-1.6B} across all datasets, demonstrating superior capabilities in generating relevant and faithful answers. Its strength was most evident in reasoning-heavy and semantically dense texts, confirming its suitability for Arabic generation tasks within RAG pipelines.

\begin{table*}[t]
    \centering
    \caption{Overall RAG performance scores for \texttt{StableLM} and \texttt{Aya-8B} across datasets. The best-performing score per dataset is shown in \textbf{bold}.}
    \label{tab:llm_results}
    \begin{tabular}{l|cc}
        \hline
        \textbf{Dataset} & \textbf{StableLM (Overall)} & \textbf{Aya-8B (Overall)} \\
        \hline
        ARCD           & 74  & \textbf{76.79} \\
        ArSQUAD        & 53.08  & \textbf{53.88} \\
        SaudiWiki      & 81.77& \textbf{88.81}\\
        QA4MRE         & 55.07& \textbf{61.8}\\
        Quran Tafseer  & 71.24& \textbf{78.39}\\
        Hindawi Books  & 67.3& \textbf{75.95}\\
        \hline
        \textbf{Average} & 67.08& \textbf{72.06}\\
        \hline
    \end{tabular}
\end{table*}

\section{Threats to Validity}
\label{sec:threats}

\noindent \textbf{Internal Validity:}  
Potential internal threats arise from pipeline implementation and metric computation. While we used standardized tools (LangChain, RAGAS, FAISS) and verified component behavior through controlled experiments, minor discrepancies in chunking boundaries or embedding normalization may have affected retrieval outcomes. We mitigated this by using fixed random seeds and consistent configurations across all experiments.

\noindent \textbf{Construct Validity:}  
Our evaluation depends on automated metrics provided by RAGAS, which may not fully capture the nuanced quality of natural language generation in Arabic. In particular, faithfulness and answer relevancy are estimated without human judgment. Although RAGAS is a widely accepted evaluation framework, future work should incorporate manual verification to complement automated scores.

\noindent \textbf{External Validity:}  
The generalizability of our findings may be limited by the datasets and models selected. While we evaluated six diverse Arabic datasets and tested both multilingual and Arabic-specific models, these corpora may not represent all domains or dialects. Additionally, the performance of the pipeline may vary in real-world applications involving noisy or low-resource data.

\section{Conclusion}
\label{sec:conclusion}

In this study, we conducted a comprehensive empirical evaluation of core components in Retrieval-Augmented Generation (RAG) pipelines tailored for Arabic language tasks. Through systematic experimentation across six diverse Arabic datasets, we assessed the impact of chunking strategies, embedding models, reranking mechanisms, and large language models on overall RAG performance using the RAGAS evaluation framework.

Our findings show that sentence-aware chunking provides the most consistent improvements in both retrieval quality and generation outcomes, particularly in reasoning-heavy and unstructured datasets. Among the embedding models, BGE-m3 and multilingual-e5-large outperformed both Arabic-specific and other multilingual alternatives, highlighting the benefits of large, contrastively trained models. The integration of the BGE reranker (v2-m3) significantly boosted faithfulness and relevancy in complex tasks, while Aya-8B demonstrated superior generation performance over StableLM-1.6B, especially in semantically dense domains.

These results offer practical guidance for building robust Arabic RAG systems and emphasize the importance of carefully selecting and tuning each pipeline component for optimal performance. The insights presented here also highlight the maturity of current multilingual technologies in supporting Arabic NLP, while pointing to opportunities for improvement in embedding and generation models specifically optimized for Arabic.

Future work will explore hybrid chunking strategies that combine sentence boundaries with semantic cohesion, as well as domain-specific rerankers tailored for legal, medical, or religious content. Additionally, human evaluation of generated answers and broader coverage of Arabic dialects will further strengthen the generalizability and reliability of RAG systems in real-world applications.

\backmatter

\section*{Declarations}

\bmhead{Conflict of Interest} The author declares no conflicts of interest.

\bmhead{Acknoweldgment} The author declares no conflicts of interest.

\end{document}